\documentstyle[12pt]{article}
\def\IR{\relax{\rm I\kern -.18em R}}

\begin{document}
\renewcommand{\theequation}{\arabic{section}.\arabic{equation}}
\title{ AN OPERATOR VALUED EXTENSION OF THE SUPER KdV EQUATIONS}
\author{ \Large S. Andrea*, A. Restuccia**, A. Sotomayor*}
\maketitle{\centerline {*Departamento de Matem\'{a}ticas,}
\maketitle{\centerline{**Departamento de F\'{\i}sica}}
\maketitle{\centerline{Universidad Sim\'on Bol\'{\i}var}}
\maketitle{\centerline{Apdo.89000, Caracas 1080 A, Venezuela}}
\maketitle{\centerline{Fax:+58-2-9063601 }}
\maketitle{\centerline{e-mail: sandrea@usb.ve, arestu@usb.ve
sotomayo@fis.usb.ve, }}
\begin{abstract}
An extension of the Super KdV integrable system in terms of
operator valued functions is obtained. Following the ideas of
Gardner, a general algebraic approach for finding the infinitely
many conserved quantities of integrable systems is presented. The
approach is applied to the above described system and infinitely
many conserved quantities are constructed. In a particular case
they reduce to the corresponding conserved quantities of Super
KdV.
\end{abstract}

\section{Introduction}
It was shown in [Mathieu] that among the one parameter
supersymmetric extensions of the KdV equation there is a special
system that has an infinite number of conservation laws. This
system is equivalent to the Super KdV equations obtained in [Manin
et al.] by reduction from the Super-Kadomtsev-Petviashvili
hierarchy. $N=2$ Supersymmetric extension of the KdV equations
have also been obtained in [Labelle et al.] [Krivonos et al.]
[Delduc et al.].

The supersymmetric extension of the KdV equation is a system of
coupled equations for a commuting and an anti-commuting field. The
commuting field $u (x,t)$ takes values in the even part of a
Grassmann algebra ${\cal G}$ while the anti-commuting field
$\xi(x,t)$ takes values on the odd part of ${\cal G}$. The
explicit form of the supersymmetric extension with an infinite
number of conservation laws is [Mathieu]
\newpage
\begin{eqnarray}u_t&=&-u'''+ 6 uu'-3\xi\xi'',\\
\xi_t&=&-\xi'''+3(\xi u)'.\end{eqnarray}

If we use, as a particular case, the four dimensional Grassman
algebra with generators $e_1$ and $e_2$, and express
\begin{eqnarray}u(x,t)&=&u_0(x,t)+u_{12}(x,t)e_1e_2,\\
\xi(x,t)&=&\xi_1(x,t)e_1+\xi_2(x,t)e_2,\end{eqnarray}

the system (1)may be reformulated as a coupled system in terms of
real or complex fields $u_0,u_{12},\xi_1\xi_2$ in the following
way
\begin{eqnarray}
u_{0 t}&=&-u_0'''+6u_0 u'_0,\\
 \xi_{1 t}&=&-\xi'''_1+3(\xi_1u_0)',\\
 \xi_{2 t}&=&-\xi'''_2+3(\xi_2 u_0)',\\
 u_{12t}&=&-u'''_{12}+6(u_0
 u_{12})'-3(\xi_1\xi''_2-\xi_2\xi''_1).
\end{eqnarray}
(1.5) is exactly the KdV equation. (1.6)and (1.7) are linear
homogeneous in $\xi_1$ and $\xi_2$ respectively. (1.8) is also
linear in $u_{12}$ but contains a source in terms of $\xi_1$ and
$\xi_2$. The Super KdV system (1) is not the only integrable
extension of the KdV equation constructed using a single
anti-commuting field. Another such system is the one proposed in
[Kupershmidt]:

\begin{eqnarray}
u_t&=&-u'''+6 u u'-3\xi\xi'',\\ \xi_t&=&-4\xi'''+6\xi'u+3\xi u'.
\end{eqnarray}
This system also has an infinite number of conservation laws. Its
expansion in the particular Grassmann algebra generated by $e_1$
and $e_2$ gives again for the $u_0$ the KdV equation . In the case
of $N=2$ supersymmetric systems, the equation for $u_0$ is
modified by nonlinear terms coming from the other even field of
the $N=2$ superfield. But again, there is no contribution to this
equation from the odd fields. In this work we present an extension
of the Super KdV system in terms of coupled system of partial
differential equations which yields a nonlinear modification for
the KdV equation with an interacting term constructed from complex
spinors. Our formulation will be in terms of operator valued
functions which for a particular case reduces to the
supersymmetric algebra. The final form of the nonlinear system
will be in terms of commuting real or complex functions in
contrast with (1.1, 1.2). There is also a physical motivation for
our program. The classical formulation of supersymmetric field
theory is always in terms of commuting and anti-commuting fields
which after quantization yield a field theory in terms of bosonic
and fermionic operators. However, once we have the quantum field
theory we may consider the mean values of the bosonic and
fermionic operators. We then obtain real scalar or vector fields
from the bosonic sector, while complex spinors come from the
fermionic sector. That is, for many applications the quantum
theory may be analyzed in terms of real fields and complex
spinors, without introducing the commuting and anti-commuting
fields of the classical formulation. For example all the analysis
of the spontaneous breaking of supersymmetry is formulated in in
those terms . It is then natural to ask if there is an extension
of the KdV equation in terms of a real field $u(x,t)$ and a
complex spinor $\xi(x,t)$. One particular case of the general
approach we will discuss is the following system

\begin{eqnarray}
u_t&=&-u'''+6u
u'-3(\varphi_1\varphi''_2-\varphi''_1\varphi_2)\nonumber\\
\varphi_{1t}&=&-\varphi'''_1+3(\varphi_1u)'\nonumber\\
\varphi_2t&=&-\varphi'''_2+3(\varphi_2u)'
\end{eqnarray}
where we consider only one complex field
$\Psi(x,t)=\varphi_1(x,t)+i\varphi_2(x,t)$. System (1.11) should
be compared with (1.5 to 1.8) where the anti-commuting fields in
(1.1 and 1.2) have been explicitly expanded in terms of a basis of
a Grassmann algebra with real or complex coefficients.

There is also another general, but may be indirect, motivation to
our program. It has been recently recognized in the formulation of
M-theory as a matrix model [IKKT] that the construction of the
theory in terms of elements of a complex Lie algebra equipped with
an invariant bilinear inner product is not only directly related
to superstring theory but also contains Yang-Mills theory, for a
suitable election of the Lie Algebra. It is well known the
relation between self-dual Yang-Mills equations and integrable
systems, in particular to the KdV hierarchy. It is then natural to
ask for an integrable generalization of the self-dual Yang-Mills
equations in terms of operator valued geometrical objects related
to the IKKT formulation of M-theory and its relation to KdV
hierarchy. Our contribution in this paper may be a first step in
this construction.

In Section 2 we present following the ideas of Gardner, a general
algebraic approach for obtaining the infinitely many conserved
quantities of certain integrable systems. We consider as a
particular case of our analysis the conserved quantities
associated with the system (1.11). In section 3 we present a
generalization of the Super KdV system in terms of operator valued
functions. We apply the general approach developed in Section 2 to
prove the existence of the infinitely many operator valued,
conserved quantities of the integrable system.

In Section 4 we discuss our conclusions.

\section {An Algebraic Approach to Nonlinear Systems and Their Conservation Laws}

\noindent In order to study nonlinear equations and their
conservation laws, one can start by choosing a ring $V$ of
infinitely differentiable functions $\IR \rightarrow \IR$ which
satisfies ${d\over dx}V\subset V$. A system of equations gives a
flow in the manifold ${\cal {M}}=V \times V \times ...\times V$,
whose general element is written $u(x)= (u_1(x), u_2(x),
...u_n(x))$. The general model for formulas involving sums of
products of derivatives of the $u_p(x)$ is a polynomial

$$f(a_{10}, a_{11},a_{20},...)$$ in a finite number of the
commuting symbols $ a_{pm}$ with $1\leq p\leq n, \leq 0\leq m<
\infty$. Then, replacing $a_{pm}$ by ${\left( {d\over dx}\right)^m
u_p(x) }$, an element $u$ of the manifold $\cal{M}$ is taken to

$$f(u)= f(u_1(x),u_1^\prime (x), u_2(x), ....),$$ an element of
$V$ whose derivative is given by

$${d\over dx}f(u)=(Df)(u)$$ $$ D={\sum_{p=1}^n} \
{\sum_{m=0}^\infty}a_{p,m+1}{\partial\over\partial a_{pm} }$$

\noindent The commutative ring ${\cal{A}}$ consisting of all such
polynomials $f$, together with the derivation
$D:{\cal{A}}\rightarrow{\cal{A}}$, may be called the free
derivation ring on $n$ generators.

\noindent The algebra $Op{\cal{A}}$, on the other hand, consists
of the linear operators $L: {\cal{A}}\rightarrow{\cal{A}}$ which
have the form $L= \Sigma_{m=0}^N l_m D^m$ with $l_m\in{\cal{A}}$.
The standard operator transpose anti-involution
$Op{\cal{A}}\rightarrow Op{\cal{A}}$ sends $L$ to $L^*$, where
$L^* f=\Sigma_{m=0}^N (-l)^m D^m(l_m f)$. Then $fLg$ and $gL^*f$
differ by an element of $\cal{DA}$ for all $f, g \in {\cal{A}}$.

\noindent Given $u\in{\cal{M}}$ and $L\in Op{\cal{A}}$, the
function substitution operation replaces

$$L(a, D)= \sum l_m (a_{10}, a_{11}, a_{20}, ...)D^m$$ by
$$L\left( u, {d\over dx}\right)=\sum l_m(u_1(x), u^\prime_1(x),
u_2(x) ...)\left(d\over dx\right)^m$$ a variable coefficient
differential operator

$$L\left(u, {d\over dx}\right): V\rightarrow V$$

\noindent In order to see how ODE systems are related by internal
substitutions, the following constructions are performed in
${\cal{A}}$ and in $Op{\cal{A}}$.

\noindent 1. A ring homomorphism $S:{\cal{A}}\rightarrow
{\cal{A}}$ which commutes with $D:{\cal{A}}\rightarrow {\cal{A}}$
is completely determined by its values on the generators $a_{p0},
1\leq p \leq n$. Conversely, if $b=(b_1, ..., b_n)$ is any $n$-
tuple of elements of $\cal{A}$, there is a unique ring
homomorphism ${\cal{S}}_b:{\cal{A}}\rightarrow{\cal{A}}$ which
commutes with $D$ and satisfies ${\cal{S}}_b(a_{p0})=b_p, 1\leq
p\leq n$. The effect of ${\cal{S}}_b$ on $f(a_{10}, a_{11},
a_{20}...)$ is to replace $a_{pm}$ by $(D^m b_p)(a_{10}, a_{11},
a_{20},... )$.

\noindent The transformation ${\cal{S}}_b$ also sends
$Op{\cal{A}}\rightarrow Op{\cal{A}}$. If $L=\Sigma l_m D^m$ then
${\cal{S}}_b L=\Sigma ({\cal{S}}_b l_m)D^m$. Again, as in
${\cal{A}}$, ${\cal{S}}_b$ preserves sums and products.

\noindent 2. In the special case $b_p= a_p + t e_p (a_{10},
a_{11}, a_{20}...)$ with $t \ \in \IR$ and arbitrary $e_p \ \in \
{\cal{A}} $, one can take the derivative at $t=0$ of ${\cal{S}}_b
f$. The result is

$$ \sum_{p=1}^n\sum_{m=0}^\infty {\partial f\over \partial a_{pm}
}D^m e_p=\sum_{p=1}^n \partial_p f (a, D)e_p$$ where, for $1\leq p
\leq n,$ the Fr\'{e}chet derivative operator $\partial_p f\in
{\cal{O}}p{\cal{A}}$ is defined to be

$$ \partial_p f (a, D)=\sum_{m=0}^\infty \ {\partial
f\over\partial a_{pm} }D^m$$

Taking the derivative of the equation ${\cal S}_b Df=D{\cal S}_b
f$, one sees that the Fr\'{e}chet derivative operators of $Df$ are
given by

$$\partial_p(Df)=D(\partial_p f). $$

When $\partial_p f$ is applied to $a_{p1}=Da_{p0}\in{\cal{A}}$,
one gets

$$\sum_{p=1}^n(\partial_p f)a_{p1}=Df.$$

3. Returning to the case of general $b$ one can verify the chain
rule

$$\partial_p {\cal{S}}_b f=\sum_{q=1}^n({\cal{S}}_b\partial_q
f)(\partial_p b_q).$$

The first step is to apply the usual chain rule, obtaining

$${\partial\over\partial a_{p m}}{\cal{S}}_b f=\sum_{q=1}^n
\sum_{r=0}^\infty \left({\partial f \over\partial a_{q
r}}\mid^{a_{qr=D^r b_q} }\right){\partial\over\partial a_{p
m}}(D^r b_q)$$.

Multiplying on the right by $D^m$ and summing over $0\leq m
<\infty$ one gets

$$\partial_p {\cal{S}}_b f=\sum_{q=1}^n \sum_{r=0}^\infty
 \left({\partial f \over\partial a_{q
r}}\mid^{a_{qr=D^r b_q} }\right)(\partial_p D^r b_q)=\sum_{q=1}^n
({\cal{S}}_b
\partial_q f)(\partial_p b_q)$$.

This completes the proof.

4. Nonlinear ODE systems are given by n-tuples $g=(g_1,...,g_n)$
of elements of ${\cal{A}}$. The unknown functions $v_p(x,t)$ are
required to satisfy

$${\partial\over\partial t}v_p (x,t)=g_p (v_1(x,t),
{\partial\over\partial x} v_1(x,t), v_2(x,t),...).$$

It then follows that

$${\partial\over\partial t}\left({\partial \over\partial x}
\right)^m v_p (x,t)=(D^m g_p)(v_1(x,t), ...).$$

5. Given $h\in{\cal{A}}$, The formula

$$ H(v)= \int h (v)d x$$ defines a (nonlinear) functional $H:{\cal
M }\rightarrow \IR$ when, for example, $v$ is the space of
infinitely differentiable $2\pi$-periodic functions and the
integral is taken from 0 to $2\pi$.

If $v(x,t)$ satisfies ${\partial\over\partial t}v=g(v)$ then the
derivative of H along the solution is given by

$${d\over d t}H=\int \sum_{p=1}^\infty \sum_{m=0}^\infty {\partial
h\over \partial a_{pm}}(v) \left({d\over d
t}\left({\partial\over\partial x}\right)^m v_p(x, t)\right)dx$$

$$ =\int \sum_{p=1}^m \partial_p h \left(v, {d \over d x}
\right)g_p(v)d x$$.

For $H$ to be a conservation law for ${\dot {v}}=g(v)$ it suffices
that

$$\sum_{p=1}^m(\partial_p h)g_p\in{\cal{DA}}. $$

6. Given $b=(b_1,..., b_n)$, there arises the transformation
$v=b(u)$ of ${\cal{M}}$ into itself, where

$$v_p(x)=b_p(u_1(x), u_1'(x), u_2(x),...). $$

If $u=u(x,t)$ satisfies the ODE system ${\dot{u}}=f(u)$, then
$v=b(u)$ satisfies

$$ {\partial\over\partial t}v_p
(x,t)=\sum_{q=1}^n\sum_{m=0}^\infty{\partial b_p\over\partial
a_{qm}}(u)\left({\partial\over\partial
t}\left({\partial\over\partial x}\right)^m u_q(x,t)\right)$$

$$=\sum_{q=1}^n \partial_q b_p\left(u,{\partial\over\partial x
}\right)f_q(u). $$

On the other hand

\begin{eqnarray}
 g_p(v)&=&g_p(b(u))\nonumber  \\
&=&({\cal S}_b g_p)(u). \nonumber\end{eqnarray}

Therefore, in order for $v=b(u)$ to be a "Miura transformation"
taking solutions of ${\dot{u}}=f(u)$ to solutions of ${\dot
{v}}=g(v)$, the equations

$$\sum_{q=1}^n (\partial_q b_p)f_q={\cal S}_b g_p$$ should hold in
$\cal A$.

7. We should expect the pullback of a conserved quantity to be a
conserved quantity. Suppose that $h\in{\cal A}$ gives a conserved
quantity for ${\dot {v}}=g(v)$, and that $v=b(u)$ is a Miura
Transformation to ${\dot {v}}=g(v)$ from ${\dot{u}}=f(u)$.

Then the b-pullback of $h$, that is to say ${\cal S}_b h$, has
Fr\'{e}chet derivative operators given by

\[ \partial_p{\cal S}_p h=\sum_{q=1}^n({\cal S}_b \partial_q h)(\partial_p
b_q)\].

After applying this equation to $f_p\in{\cal A}$ we get

\[ \sum_p(\partial_p {\cal S}_b h)f_p=\sum_{p,q}({\cal S}_b\partial_q h)
(\partial_p b_q)f_p\]
\[=\sum_q ({\cal S}_b\partial_q h)({\cal S}_b g_q)\]
because $v=g(u)$ is a Miura transformation. Then, ${\cal S}_b$
being a ring homomorphism, \begin{eqnarray} \sum_p(\partial_p{\cal
S}_b h)f_p&=&{\cal S}_b\left(\sum_q(\partial_q h)g_q\right
)\nonumber
\\  &=&{\cal S}_b {\cal D}e \nonumber
\end{eqnarray}
for some $e\in{\cal A}$, because $h$ gives a conserved quantity
for ${\dot{v}}=g(v)$. Since ${\cal S}_b$commutes with ${\cal D}$,
we may conclude that ${\cal S}_b h$ gives a conserved quantity for
${\dot{u}}=f(u)$.

\section{Conservation Laws for a KdV System}
This theory will now be applied to a generalization of the $KdV$
equation. The usual version corresponds to the element
$-a_1{'''}+6 a_1 a'_1$ in the free derivation ring on one
generator,where the double suffix notation is shortened to
$a_p=a_{p0},a'_p=a_{p1},a''_p=a_{p2}, etc.$, for small values of
$m$ in $a_{pm}$.

For the extended $KdV$ we go from $n=1$ to $n=3$ and set

\begin{eqnarray}
g_1&=&(-a''_1+3a_1^2+3[a_2,a_3])'\nonumber\\ g_2&=&(-a_2''+3a_1
a_2)'\nonumber\\g_3&=&(-a''_3+3a_1 a_3)'\nonumber
\end{eqnarray}
in which $[a_p, a_q]=a'_p a_q-a_p a'_q$ in general.

This system, to be called ${\dot{v}}=g(v)$, is subjected to the
transformation ${\cal S}_b:{\cal A}\rightarrow{\cal A} $ where

\[\pmatrix{b_1&\cr b_2&\cr b_3}=\pmatrix{a_1&\cr a_2&\cr a_3}+\varepsilon
\pmatrix{a'_1&\cr a'_2 &\cr
a'_3}+\varepsilon^2\pmatrix{a_1^2+[a_2,a_3]&\cr a_1 a_2&\cr a_1
a_3}\] and $\varepsilon$ is any real number. When $\varepsilon=0,
{\cal S}_b$ reduces to the identity map. Since the $g_p$ are
quadratic one gets

\[{\cal S}_b g=\sum_{i=0}^4\varepsilon^i{\cal C}_i\]
in which $g$ is the column vector $(g_1,g_2,g_3)^T$. Evidently
${\cal C}_0=g$ and ${\cal C}_1={\cal D}g$, the latter because
${d\over d \varepsilon}|^{\varepsilon=0}{\cal S}_b
f=\sum_p(\partial_p f )a'_p={\cal D}f$ for any $f\in{\cal A}$.

The last two columns are

\[{\cal C}_3=\pmatrix{(2a_1^3+3a_1[a_2,a_3])'
\cr 3a_1(a_1 a_2)'+3a'_2[a_2,a_3]\cr 3a_1(a_1
a_3)'+3a'_3[a_2,a_3]}'\] and

\[{\cal C}_4=\pmatrix{3a_1^4+9a_1^2[a_2,a_3]+3[a_2,a_3]^2
\cr3a_1^3a_2+3a_1a_2[a_2,a_3] \cr3a_1^3a_3+3a_1a_3[a_2,a_3]}'\]

Turning now to the Fr\'{e}chet derivative operators of the three
components of $b$, we find the $3\times3$ matrix of elements of
${\cal O }_p{\cal A}$ which is given by

\[\partial_q b_p=I+\varepsilon {\cal D}+\varepsilon^2 B_2\] in
which

\[B_2=\pmatrix{2a_1&[*,a_3]&[a_2,*]\cr a_2&a_1&0\cr a_3&0&a_1}.\]

In order for a system $\dot{u}=f(u)$ to be sent to
${\dot{v}}=g(v)$ by $v=b(u)$, the coefficients $f_p$ must satisfy

\[{\cal S}_b g_p=\sum_{q=1}^3(\partial_q b_p)f_q\].

The corresponding equation of column vectors, expanded in powers
of $\varepsilon$, is

\[\sum_{i=0}^4\varepsilon^i{\cal C}_i=(I+\varepsilon {\cal D}+\varepsilon^2 B_2)
\sum_{j=0}^\infty \varepsilon^j F_j.\]

All the column vectors $F_j$ are determined recursively; in
particular $F_0={\cal C}_0=g, F_1=0,$and

\[F_2=\pmatrix{(2a_1^3+3a_1[a_2,a_3])'
\cr 3a_1(a_1 a_2)'+3a'_2[a_2,a_3]\cr 3a_1(a_1
a_3)'+3a'_3[a_2,a_3]}\]

It now turns out that $F_j=0$ for $j\geq3$, permitting the fourth
degree polynomial ${\cal S}_b g_p$ to be written as the product of
$(\partial_q b_p)$ by $f_q$ with $f=(f_1,f_2,f_3)^T,
f=g+\varepsilon^2F_2$, that is, as a product of two quadratic
polynomials in $\varepsilon$. This completes the construction of
$\dot{u}=f(u)$, a system sent to ${\dot{v}}=g(v)$ by $v=b(u)$.

Conservation laws for $\dot{u}=f(u)$ are given by $h\in{\cal A}$
satisfying $\sum_{p=1}(\partial_p h)f_p\in {\cal DA}$. Since
$\partial_pa_q=\delta(p,q)$ in general, the choice $h=a_1$ gives
just $f_1$, which by the construction of $f$ is always in ${\cal
DA}$. Therefore $h=a_1$ is indeed a conserved quantity for
${\dot{u}}=f(u)$.

If the ring homomorphism ${\cal S}_b:{\cal A}\rightarrow {\cal A}$
could be inverted, the pullback of $a_1$ would give a conservation
law for ${\dot{v}}=g(v)$, for all $\varepsilon$. But it can indeed
be inverted if we embed ${\cal A}$ within ${\cal
A[[\varepsilon]]}=B$, whose elements are the formal power series
in $\varepsilon$ with coefficients in ${\cal A}$. The given
elements $b_p\in B, 1\leq p\leq 3$, define a unique ring
homomorphism ${\cal S}_b:{\cal A}\rightarrow B$ which sends
$a_{p0}$ to $b_p$ and commutes with ${\cal D}$. Obviously it
extends to ${\cal S}_b:B\rightarrow B$.

But when $\varepsilon=0$ the element $b_p\in B$ reduces to the
element $a_p\in {\cal A}$. Therefore, for any $h(a_{10},
a_{11},a_{20}...)\in{\cal A}$, ${\cal S}_b h$ will have the form

\[{\cal S}_b h=h+\varepsilon h_1+\varepsilon^2h_2+...\]
for some $h_1, h_2,...\varepsilon{\cal A}$.

Within $B$ one has the ideals ${\cal B}_k=\epsilon^k{\cal B}$ and
the filtration $B\supset B_1\supset B_2\supset...$ . The preceding
observation shows that ${\cal S}_b$ sends each $B_k$ into itself,
and reduces to the identity map in each quotient space
$B_k/B_{k+1}$. This shows that, given any $f_0, f_1...$ in $\cal
A$, the equation

\[{\cal S}_b\left(\sum_{k=0}^\infty\varepsilon^k g_k\right)=
\sum_{k=0}^\infty\varepsilon^k f_k\] can be solved recursively for
$g_0,g_1,...\varepsilon{\cal A}.$ Therefore ${\cal
S}_b:B\rightarrow B$ is an isomorphism.

In order to see the recursion algorithm more clearly we work in
the space ${\cal A}\bigoplus {\cal A}\bigoplus{\cal A}$ of column
vectors $e=(e_1,e_2,e_3)^T$. Then

\[b=a+\varepsilon a'+\varepsilon^2<a,a>\] where in general

\[<f,g>=(f_1g_1+[f_2,g_3],f_1g_2,f_1g_3)^T\],
a bilinear map of ${\cal A}\bigoplus {\cal A}\bigoplus{\cal A}$
into itself. We ask that column vectors ${\cal C}_0,{\cal
C}_1,{\cal C}_2,...$ be determined in such a way that

\[{\cal C}={\cal C}_0+\varepsilon{\cal C}_1+\varepsilon^2{\cal C}_2+...\]
satisfies

\[a={\cal C}+\varepsilon{\cal C}'+\varepsilon^2<{\cal C},{\cal C}>.\]
That is, the ring homomorphism ${\cal S}_{{\cal C}}:B\rightarrow
B$ should send $b_p$ to $a_p$ for $1\leq p\leq 3.$

After equating the coefficients of corresponding powers of
$\varepsilon$ one finds the recursion relation

\[0={\cal C}_{m+2}+{\cal C}'_{m+1}+<{\cal C}_m,{\cal C}_0>+<{\cal C}_{m-1},{\cal C}_1>
+...+<{\cal C}_0,{\cal C}_m>\]

The values for $0\leq m\leq 4$ are

\begin{eqnarray}{\cal C}_0&=&a\nonumber \\{\cal C}_1&=&-a' \nonumber \\
{\cal C}_2&=&a''-<a, a>\nonumber \\ {\cal
C}_3&=&-a'''+2<a,a>'\nonumber \\ {\cal
C}_4&=&a''''-3<a,a>''+<a',a'>+<<a,a>,a>+<a<a,a>>.\nonumber
\end{eqnarray}

By construction , the transformation ${\cal S}_{\cal C}$ takes
${\dot{v}}=g(v)$ to $\dot{u}=f(u)$. The latter equation has $a_1$
as a conserved quantity. Its pullback is the top entry in the
column ${\cal C}={\cal C}(\varepsilon)$.

As this is true for all $\varepsilon$, we conclude that the top
entries of all the columns ${\cal C}_m$ give conserved quantities
for the extended $KdV$ equation ${\dot{v}}=g(v)$.

Simplifying where possible by subtracting elements of ${\cal DA}$
or by changing signs, the first three nontrivial functionals
${\cal M}\rightarrow R$ are

\[H_m(v)=\int h_m(v_1(x),v'_1(x),v_2(x),...)dx\]

with

\begin{eqnarray}
h_0&=& a_1\nonumber\\
h_2&=&(a_1)^2+a'_2a_3-a_2a'_3\nonumber\\h_4&=&2(a_1)^3+(a'_1)^2+a''_2a'_3-a'_2a''_3
+4a_1(a'_2a_3-a_2a'_3).\nonumber
\end{eqnarray}

Evidently, the extended $KdV$ equation has infinitely many
conservation laws.

\section{ Extension of the KdV equation to operator - va-lued functions.}
\setcounter{equation}{0}
This extension, which also has infinitely
many conservation laws, includes the supersymmetric KdV as a
special case, as well as the extension seen in the preceding
section.

That extension is a system of equations in the three real-valued
functions $v_p(x,t)$, $1\leq p\leq3$. It is recast in operator
form by writing
\begin{eqnarray} {\cal P}&=&v_1 (x, t)I\nonumber \\
{\cal Q}&=& v_2 (x,t)E_2+v_3(x,t)E_3\nonumber
\end{eqnarray}
where $E_2$ and $E_3$ are linear operators in some space which
satisfy $[E_2,E_3]=E_2 E_3-E_3 E_2=I$. The quantity $v_2'v_3-v_2
v_3'$, which was denoted before by $[v_2,v_3]$ will now appear as
the coefficient of $I$ in the usual commutator $[{\cal Q}', {\cal
Q }]$. Thus we have the equivalent system of operator differential
equations

\begin{eqnarray}{\cal P}_t&=& -{\cal P}'''+6 {\cal P}{\cal P}'+ 3[{\cal
Q}'',{\cal Q}]\\ \nonumber{\cal Q}_t&=&-{\cal Q}'''+3({\cal
P}{\cal Q})'.
\end{eqnarray}

Further, setting $p=u_1(x,t)I$ and $q=u_2(x,t)E_2+u_3(x,t)E_3$,
the Gardner transformation takes the form

\begin{eqnarray}{\cal P}&=&p+\epsilon p'+\epsilon^2(p^2+[q',q])\\
\nonumber{\cal Q}&=&q+\epsilon q' +\epsilon^2pq.
\end{eqnarray}
and the modified KdV equation $\dot{u}=f(u)$ takes the form

\begin{eqnarray}p_t&=&(-p''+3p^2+3[q',q])'+\epsilon^2(2p^3+3p[q',q])' \\ \nonumber
q_t&=&(-q''+3pq)'+\epsilon^2 3(p^2q'+p p'q+q'[q',q]).
\end{eqnarray}

It was shown in the preceding section that the Gardner
transformation takes a solution $p,q$ of the latter system to a
solution ${\cal P},{\cal Q}$ of the former.

But more generally one can suppose that $P$ and $p$ have values in
${\cal P}$, a commutative algebra with unit, of operators acting
in some vector space. Then, if ${\cal Q}$ is a linear space of
operators satisfying

\[[{\cal Q}, {\cal P}]=0 \] \begin{equation}{\cal Q}{\cal
P}\subset {\cal Q}\end{equation} \[[{\cal Q},{\cal Q}]\subset
{\cal P}\] then ${\cal Q}$ and ${q}$ can take their values in
${\cal Q}$.

For the specific choice ${\cal P}=\{\alpha I\},{\cal Q}={\{\beta_2
E_2+\beta_3 E_3\}}$ just considered, we have observed that the
Gardner transformation (4.2) takes solutions of (4.3) to solutions
of (4.1), this being no more than a restatement of the results of
the preceding section.

However, upon reexamining the calculations in that section, one
sees that they remain valid not just for the ${\cal P}, {\cal Q}$
just considered but for any ${\cal P}$ and ${\cal Q}$ satisfying
(4.4).

Therefore the conservation laws of (4.1) and of (4.3) are
interrelated by the Gardner transformation (4.2) and its inverse,
just as before; in particular (4.1) has infinitely many
conservation laws.

After simplifying by crossing out derivatives in $x$, the first
four nontrivial conserved quantities for the operator - extended
KdV system are

\begin{eqnarray}
H_0&=&\int {\cal P}d x \nonumber \\ H_2&=&\int ({\cal P}^2+[{\cal
Q}', {\cal Q}])d x \nonumber \\ H_4&=&\int(2{\cal P}^3+({\cal
P}')^2+4{\cal P}[{\cal Q}',{\cal Q}]+[{\cal Q}'',{\cal Q}'])d x
\nonumber \\ H_6&=&\int (5{\cal P}^4+10{\cal P}({\cal
P}')^2+({\cal P}'')^2+15{\cal P}^2[{\cal Q}',{\cal Q}]-2{\cal
P}[{\cal Q}'',{\cal Q}']\nonumber \\ &-& 8{\cal P}[{\cal
Q}''',{\cal Q}]+3[{\cal Q}',{\cal Q}]^2+[{\cal Q}''',{\cal
Q}''])dx .
\end{eqnarray}
In each case, the conserved quantity has its values in the
operator algebra ${\cal P}$.

The foregoing theory applies to any operator spaces ${\cal
P},{\cal Q}$ having the stipulated properties. For example, the
${\cal Q}$ seen before can be enlarged to ${\cal Q}=\{
\sum_{k=1}^m(\mu_k E_k+\nu_k F_k)\}$ where
$\mu_k,\nu_k\epsilon\IR$ while $E_1,...,E_m,F_1,...,F_m$ are
linearly independent operators in some vector space, satisfying
$[E_k, F_k]=I$ but with all other commutator brackets zero. Then,
with ${\cal P}=\{ \alpha I\}$ as before, the operator - extended
KdV system becomes a system of nonlinear differential equations
for $2m+1$ functions $\alpha(x,t),\mu_k(x,t),\nu_k(x,t)$,
specifically

\begin{eqnarray}
{\partial\over\partial
t}\alpha&=&-\alpha'''+6\alpha\alpha'+3\sum_{k=1}^m(\mu_k''\nu_k-\mu_k\nu_k'')\nonumber\\
{d\over d t}\mu_k&=&-\mu_k'''+3(\alpha\mu_k)'\nonumber\\{d\over d
t}\nu_k&=&-\nu_k'''+3(\alpha\nu_k)'.
\end{eqnarray}

Another choice of ${\cal P}$ and ${\cal Q}$ leads to the
supersymmetric extension of KdV. An exterior algebra $\bigwedge$
on a finite set of generators is the direct sum
$\bigwedge=\bigwedge_0\oplus\bigwedge_1$, where $\bigwedge_0$ and
$\bigwedge_1$ consist of the linear combinations of even products,
respectively odd products, of the generators. Then ${\cal P}$ and
${\cal Q}$ are, respectively, the operators of left multiplication
in $\bigwedge$ by elements of $\bigwedge_0$ and of $\bigwedge_1$.
In this example some of the equations in the general theory to be
simplified, for example $[{\cal Q''},{\cal Q}]=2{\cal Q''}{\cal
Q}$. The change of variables ${\cal P}=u, {\cal
Q}=2^{-{1\over2}}\xi$ converts the operator - extended KdV into

\begin{eqnarray}
u_t&=&-u'''+6 u u'-3\xi\xi'' \nonumber \\ \xi_t&=&-\xi'''+3(\xi
u)',\nonumber
\end{eqnarray}
which is the supersymmetric extension of KdV given by Mathieu in
[Mathieu]. Moreover, the modified system (4.1) and the conserved
quantity $H_6$ are simplified a bit by $q'[q',q]=0$ and $[{\cal
Q'},{\cal Q}]^2=0$ in the supersymmetric case.

\section{Conclusion}
In this paper an operator valued extension of the KdV equation was
constructed. In a particular case the Super KdV equations were
recovered. A general algebraic method was developed and applied to
show that there are infinitely many conserved quantities for
certain integrable systems. When the method was applied to the
operator extension of KdV, the first few conserved quantities were
computed explicitly. The conserved quantities $H_0, H_2$ and $H_4$
of super KdV can be seen to correspond term by term with the
corresponding quantities of the operator extension. However, the
quantities $H_6, H_8...$ contain extra terms in the operator case
which reduce to zero for super KdV.

\section{References}

[Mathieu ] P. Mathieu, J. Math. Phys. 29, 2499 (1988) ; "Open
Problems for the super KdV equations", math-ph/0005007.
\newline
\newline
[Manin et al.] Yu. I. Manin and A.O. Radul, Commun. Math. Phys 98,
65 (1985.
\newline
\newline
[Kupershmidt] Kupershmidt, Phys. Lett. A102, 213 (1984).
\newline
\newline
[Gardner et al.] R. M. Miura, C. S. Gardner, and M. D. Kruskal, J.
Math. Phys. 9, 1204 (1968).
\newline
\newline
[Labelle et al.] P. Labelle and P. Mathieu, J. Math. Phys. 32, 923
(1991).
\newline
\newline
[Delduc et al.] F. Delduc, L. Gallot, and E. Ivanov, Phys. Lett.
B396, 122 (1997).
\newline
\newline
[Krivonos et al.] S. Krivonos, A. Pashnev, and Z. Popovicz, Mod.
Phys. Lett. A13, 146-35 (1998).
\newline
\newline
[IKKT] N. Ishibashi, H Kawai, Y. Kitazawa, A. Tsuliya, Nucl. Phys.
B498, 467(1997).
\end{document}